# FPGA-based Time to Digital Converter and Data Acquisition system for High Energy Tagger of KLOE-2 experiment

Lorenzo Iafolla, Alessandro Balla, Matteo Beretta, Paolo Ciambrone, Maurizio Gatta, Francesco Gonnella, Matteo Mascolo, Roberto Messi, Dario Moricciani, Domenico Riondino

*Abstract*— In order to reconstruct γγ physics events tagged with High Energy Tagger (HET) in the KLOE-2 (K LOng Experiment 2), we need to measure the Time Of Flight (TOF) of the electrons and positrons from the KLOE-2 Interaction Point (IP) to our tagging stations (11 m apart). The required resolution must be better than the bunch spacing (2.7 ns). We have developed and implemented on a Xilinx Virtex-5 FPGA a Time to Digital Converter (TDC) with 625 ps resolution (LSB) along with an embedded data acquisition system and the interface to the online FARM of KLOE-2. We will describe briefly the architecture of the TDC and of the Data AcQuisition (DAQ) system. Some more details will be provided about the zero-suppression algorithm used to reduce the data throughput.

## I. INTRODUCTION

For the KLOE upgrade (KLOE-2, [1]) we built a new couple of subdetectors: the HETs detectors [2], two position detectors for electrons and positrons dedicated to the study of the γγ physics. The detectors are made of 29 plastic scintillators coupled to 29 photomultipliers; the signals from the photomultipliers are then discriminated and shaped by the frontend electronics. In order to reconstruct the events with KLOE-2 we need a system able to:
- measure the arrival time of the discriminated signals respect to a reference signal (Fiducial);
- process the data;
- interface with KLOE-2 trigger and acquisition systems.

All these requirements led us to develop an FPGA based TDC. The technique we chose is the 4xOvesampling [2], which needs few resources and allows us to implement on the same device also the data acquisition system.

The data acquisition system is based on the peculiar necessities of interfacing it to KLOE-2 and is completely custom. A zero suppression algorithm was developed and implemented on the FPGA, too, in order to reduce the data throughput.

## II. TDC ARCHITECTURE

### A. The 4xOversampling technique

In order to have an enough big range and a good resolution we used the linear interpolation technique that is well described in literature [4],[5]. The interpolators performing the high resolution measures are based on the 4xOversampling technique. Such technique consists basically on sampling the input signal by means of 4 flip-flops synchronized with 4 clock signals out of phase of 90° one by the others (Fig. 1).

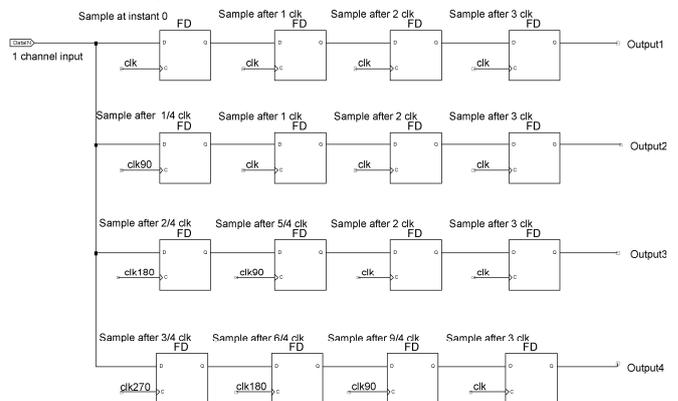

Fig. 1. The input stage for the 4xOversampling technique.

A decision logic produces a 2-bits measure of the time between the positive edge of the clock signal and the input signal. The resolution $T_0$ (the smallest time interval that can be resolved in a single measure) of the 4xOversampling interpolator is $T_{clk}/4$. We decided to use a 400 MHz clock frequency which corresponds to a 625 ps resolution and is enough for our purpose.

Manuscript received May 30, 2012. This work was supported by INFN (Italian National Institute of Nuclear Physics).

L. Iafolla, A. Balla, P. Ciambrone, M. Gatta, F. Gonnella, and D. Riondino were with the National Laboratories of Frascati (LNF) of National Institute of Nuclear Physics (INFN), via E. Fermi 40, 00044 Frascati (RM), Italy (e-mail: lorenzo.iafolla@lnf.infn.it).

M. Mascolo, R. Messi, D. Moricciani were with RM-2 Department of National Institute of Nuclear Physics (INFN), via della Ricerca Scientifica, 1, 00133 Rome, Italy.

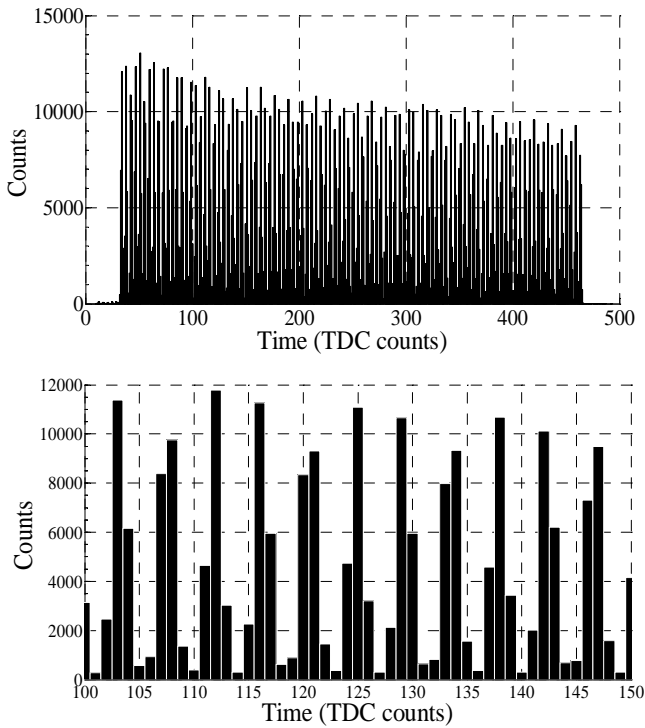

Fig. 2. Histograms of the time measures performed with the TDC fed by the HET detector installed on DAΦNE. The second plot is a detail of the first; they both show the bunches structure of the beam.

### III. TDC AND DATA ACQUISITION FOR HET TAGGERS

#### A. The 4xOversampling technique to distinguish the bunches of DAΦNE

KLOE-2 is a general purpose particle physics experiment which works at Double Annular Φ Factory for Nice Experiments (DAΦNE) accelerator.

In order to reconstruct the physics events we need to associate each detected particle to the right bunch crossing. The beam of DAΦNE has the structure of 120 bunches of which only the first 100 contains particles; the minimum time between two bunches is 2.7 ns.

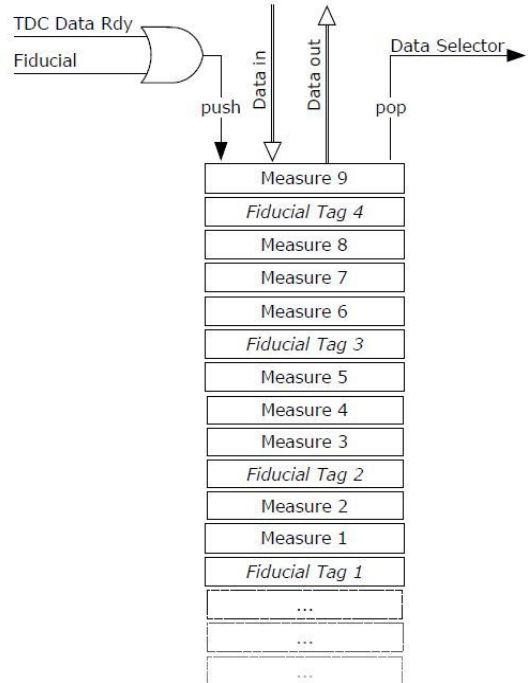

Fig. 4. The Stack buffer and an example of its content.

The association can be done by measuring the time between a signal provided by the machine (Fiducial) that identifies the first bunch, and the instant on which the particle has been detected. Fig. 2 shows the bunches structure of the beam deduced by consecutive TDC measures of the background particles. This plot can be used to associate the TDC output with the bunch crossing number.

#### B. The zero-suppression algorithm

The TDC performs measurements continuously, producing a big amount of data but only a small fraction contains valid information. In order to select and store only relevant data, the KLOE trigger signals T1 and T2 are exploited and a zero suppression algorithm is implemented to discard all the non-valid data.

The first buffer stage is a Stack memory (Fig. 4) that assures to read out only the most recent data; this is done by writing just before the first bunch a peculiar word, Fiducial Tag, to recognize the newest cycles.

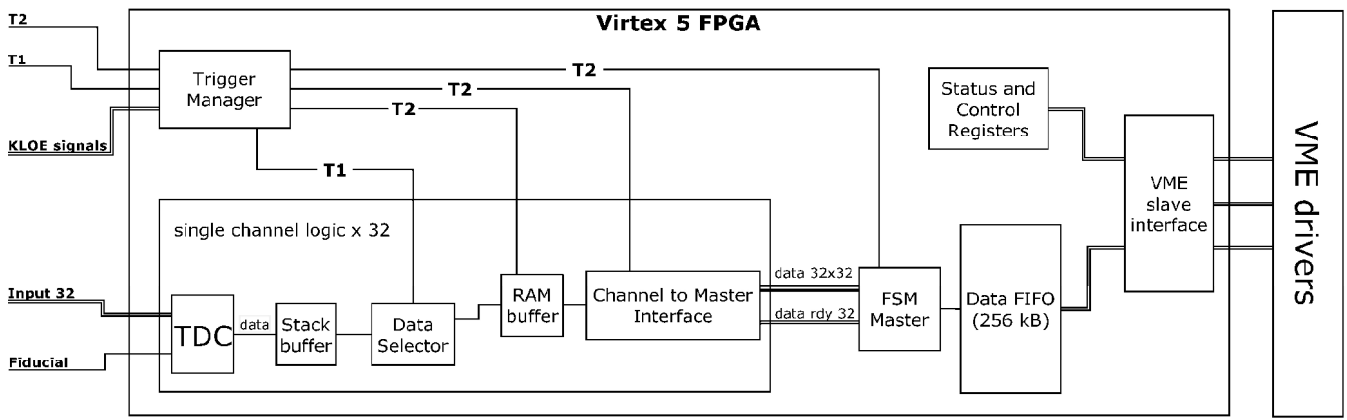

Fig. 3. Simplified scheme of the TDC and of the readout system of the HET. Some logic elements are replicated for each channel of the TDC. KLOE-2 DAQ system provides the triggers and some other control signals.

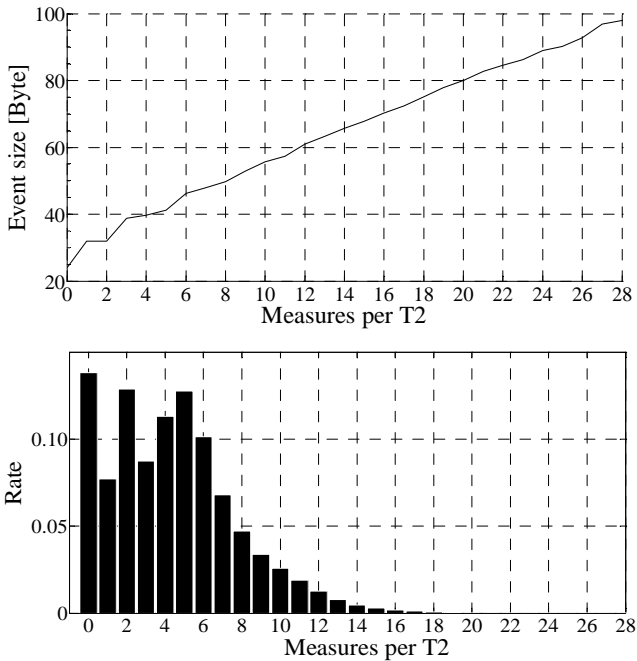

Fig. 5. The first plot is the event size vs. number of hits for event. The second plot is the distribution of the number of hits per event: it was obtained when the peak luminosity of DAΦNE was $8 \cdot 10^{31} \mathrm{cm}^{-2} \mathrm{s}^{-1}$.

In Fig. 3 a simplified scheme of the whole acquisition system is represented. The data are stored, moved and selected through the buffer stages (Stack Buffer, RAM buffer, Data FIFO) by three Finite State Machines (Data Selector, Channel to Master Interface, FSM Master). As shown in scheme, the content of the main buffer (Data FIFO) can be accessed via a A32/D64 VME64x interface. The performances of the zero-suppression algorithm can be evaluated through the relation between the event size and the number of TDC measures per T2 (first plot of Fig. 5). The average amount of data per event can be evaluated taking in account the distribution of the number of measures per T2 (second plot of Fig. 5): it is less 40 byte/T2 and is negligible in comparison with the total amount of data from the other detectors of KLOE-2 that is about 2 kByte.

*C. The Virtex-5 custom board*

We developed a general purpose VME board that hosts a FX70T Virtex-5 FPGA. Thanks to the few resources used by the TDCs, we were able to implement a complete DAQ system on the FPGA comprehensive of a 32 channels TDC (more than 29 that is the number of channels of each detector).

The board has different interfaces (VME, Ethernet, USB, RS232, Optical links) and a DDR2 RAM memory for data buffering.

In order to test and monitor the HET DAQ we also implemented some facilities for debugging and a scaler (not shown in Fig. 3) to measure the rates of detection.